\documentclass[a4paper,twocolumn,11pt,floatfix,accepted=2022-08-31]{quantumarticle}
\pdfoutput=1
\usepackage[utf8]{inputenc}
\usepackage[english]{babel}
\usepackage[T1]{fontenc}
\usepackage{amsmath}
\usepackage{hyperref}

\usepackage{tikz}
\usepackage{lipsum}

\usepackage{braket}
\usepackage{array}
\usepackage{tabularx}
\usepackage{mathtools}
\usepackage{commath}
\usepackage{bbold}
\usepackage{graphicx}
\usepackage[numbers]{natbib}

\DeclarePairedDelimiter\floor{\lfloor}{\rfloor}

\begin{document}
\title{Subradiant edge states in an atom chain with waveguide-mediated hopping}
\author{Ciaran McDonnell}
\affiliation{School of Physics and Astronomy and Centre for the Mathematics and Theoretical Physics of Quantum Non-Equilibrium Systems, The University of Nottingham, Nottingham, NG7 2RD, United Kingdom}
\orcid{0000-0002-2319-2581}
\author{Beatriz Olmos}
\affiliation{School of Physics and Astronomy and Centre for the Mathematics and Theoretical Physics of Quantum Non-Equilibrium Systems, The University of Nottingham, Nottingham, NG7 2RD, United Kingdom}
\affiliation{Institut f\"ur Theoretische Physik, Universit\"at Tübingen, Auf der Morgenstelle 14, 72076 T\"ubingen, Germany}
\orcid{0000-0002-1140-2641}

\begin{abstract}
    We analyze the topological and dynamical properties of a system formed by two chains of identical emitters coupled to a waveguide, whose guided modes induce all-to-all excitation hopping. We find that, in the single excitation limit, the bulk topological properties of the Hamiltonian that describes the coherent dynamics of the system are identical to the ones of a one-dimensional Su-Schrieffer-Heeger (SSH) model. However, due to the long-range character of the exchange interactions, we find weakening of the bulk-boundary correspondence. This is illustrated by the variation of the localization length and mass gap of the edge states encountered as we vary the lattice constant and offset between the chains. Most interestingly, we analytically identify parameter regimes where edge states arise which are fully localized to the boundaries of the chain, independently of the system size. These edge states are shown to be not only robust against positional disorder of the atoms in the chain, but also subradiant, i.e., dynamically stable even in the presence of inevitable dissipation processes, establishing the capacity of waveguide QED systems for the realization of symmetry protected topological phases.
\end{abstract}

\maketitle

\section{Introduction}

A topological phase is a form of matter which can be characterized by a non-local order parameter such as the well celebrated Berry phase \cite{berry1984,asboth2016,wen2017,Qi2011}. The properties of these topological phases are not altered by local perturbations, being inherently robust not only against local noise, but also disorder \cite{kitaev2003}. These properties have recently propelled systems that are capable of hosting such topological phases to front-runners for applications in the quantum technological context, such as in topological quantum computation \cite{chayak2008,lahtinen2017,field2018}, quantum state transfer \cite{longhi2019,dangelis2020,lang2017,litinski2017}, and quantum error correction \cite{lang2018,stricker2020}, among others. Together with solid-state systems such as resonators or metamaterials \cite{nash2015,mittal2016,stjean2017,Barik2018,ozawa2019}, cold atom setups such as Rydberg atoms \cite{weber2018,deleseleuc2019}, or dense atomic lattice gases \cite{bettles2017,perczel2017} have been shown to display symmetry-protected properties such as flat bands and robust edge states.

Systems governed by quadratic Hamiltonians and Lindbladians can be classified according to their topological properties via the so-called `tenfold way classification' \cite{ryu2010,Lieu2020,Zhou2020}. Whether or not topological phases can be hosted in the system is here determined by the dimension of the problem and the symmetries of the Hamiltonian or Lindbladian. Moreover, in these systems the bulk-boundary correspondence determines a relation between the bulk topological invariant and the amount of so-called edge states. These edge states are spatially localized on the boundaries of the system, decaying exponentially into the bulk \cite{Qi2006,Hasan2010}. Several recent works have shown (both theoretically and experimentally) that deviations from the tenfold way classification and weakening of the bulk-boundary correspondence may arise when one considers long-ranged interacting systems, e.g., with hopping constants that go beyond nearest neighbors \cite{li2014,vodola2014,vodola2015,alecce2017,lepori2017,li2018,Longhi2018,pocock2018,pocock2019,perez-gonzalez2019,Chen2020,Hsu2020,Jiao2021,Dias2022}. Many of these works consider interactions that decay as a power law of the distance between sites. This assumption is particularly well-suited for describing cold atom physical systems, e.g. ions or Rydberg atoms.

\begin{figure} 
\centering
    \includegraphics[width =0.9\columnwidth]{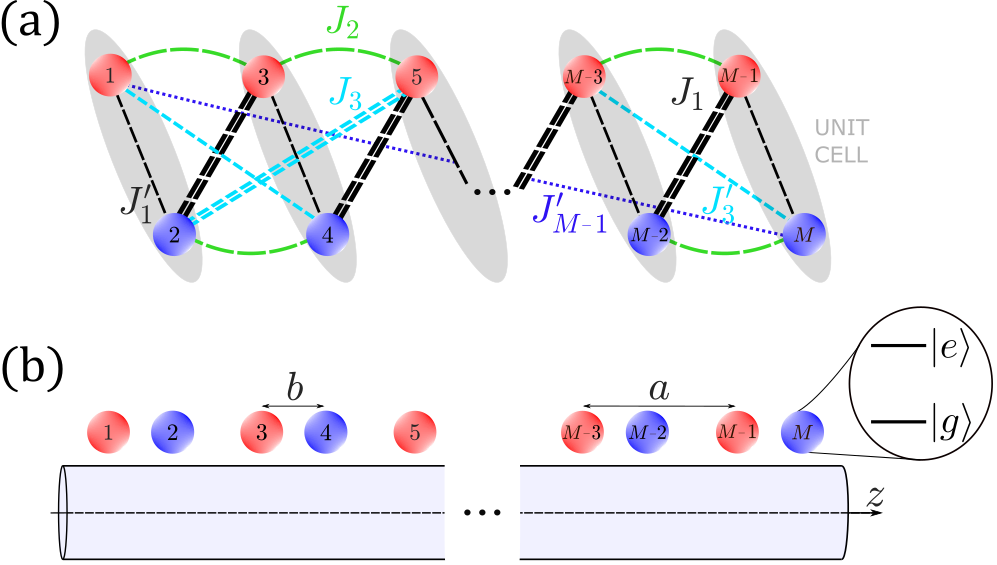}
    \caption{\textbf{SSH extended model.} \textbf{(a):} One-dimensional array of $M$ sites where each unit cell (grey) contains two sites that belong to the sublattice A and B (red and blue, respectively). Both inter and intra sublattice hopping can occur between sites $i$ and $j$. The hopping rates are defined in the main text. \textbf{(b):} The extended SSH model can be reproduced by two arrays of identical atoms (modelled as two-level systems) placed near a one-dimensional waveguide, with each array representing a sublattice. The intercell distance (lattice constant) is $a$ and the intracell atoms are offset by the distance $b$.} \label{fig:Setup}
\end{figure} 

In this work, we go a step further and analyze the topological properties of a system with infinite-range (or all-to-all) interactions. In particular, we study the topological properties and presence of edge states in a generalized version of the celebrated Su-Schrieffer–Heeger (SSH) model \cite{su1979} on a one-dimensional chain, where hopping may occur across all sites [see Fig. \ref{fig:Setup} (a)] \cite{Chen2020}. Beyond shedding light on the topological properties of systems with long-range interactions, our motivation to study such a system stems from the latest developments in quantum hybrid systems such as atoms coupled to cavities \cite{Chang2012,Chanda2021} or waveguides, such as nanofibers or photonic crystals \cite{Bello2019,ozawa2019,Kim2021}. In these systems, the translationally symmetric light modes mediate all-to-all exchange interactions among the atoms \cite{LeKien2014,LeKien2014b,Jen2022,poddubny2022}. In particular, the extended SSH model we study here describes approximately the physics of an array of two-level atoms coupled to a nanophotonic waveguide [see Fig. \ref{fig:Setup} (b)].

We find that when the system conserves chiral symmetry, the bulk Hamiltonian belongs to the same topological class (BDI) as the original SSH model with short-range hopping. However, in general here the bulk-boundary correspondence is weakened, leading to the appearance of massive edge states with non-zero energy. We uncover parameter regimes where approximate zero-energy edge states, which are robust against local disorder, arise. Interestingly, we also reveal the presence of robust edge states in the case where the chiral symmetry is broken, and hence the tenfold way classification does not predict their existence. Finally, we study the out-of-equilibrium dynamics of the system, which reveals that some of the uncovered edge states are subradiant, i.e., decay collectively emitting a photon into the waveguide in a much longer timescale than an isolated single atom would do.

\section{Physical system and model Hamiltonian}\label{sec:Model}

Let us consider a chain of $M$ identical atoms arranged in a bipartite lattice structure around a waveguide, e.g., a cylindrical optical nanofiber of refractive index $n_1$. The two atomic chains, with lattice constant $a$ and formed by $M/2$ atoms each, are placed close to the waveguide parallel to its longitudinal axis (here the $z$-axis), offset with respect to one another by a distance $b$ [see Fig. \ref{fig:Setup}(b)]. The atoms are modelled as two-level systems with ground and excited states $\ket{g_i}$ and $\ket{e_i}$ for the $i$-th atom, respectively.

Due to the coupling to the environment, the atoms incoherently emit photons and also interact with one another via dipole-dipole interactions due to the exchange of virtual photons mediated both by the guided and unguided modes of the waveguide \cite{LeKien2014,LeKien2014b,Jen2022}. Under the usual Born-Markov and secular approximations, one can describe the dynamics of this system through the quantum optical master equation
\begin{equation}\label{Master eq}
      \!\dot{\rho} \!=\! -\frac{\mathrm{i}}{\hbar}\!\left[H'\!,\rho \right] \!+\! \sum_{ij}\Gamma_{ij}\!\left(\!\sigma_{j}\rho\, \sigma^{\dag}_{i} \!-\!\frac{1}{2}\left\{\sigma_{i}^{\dag}\sigma_{j},\rho  \right\}  \!\right)\!.
\end{equation}
Here, the Hamiltonian
\begin{equation}\label{eq:Hprime}
    H'=\sum_{i\ne j}V_{ij}\sigma_{i}^{\dag}\sigma_{j}
\end{equation}
describes the coherent dipole-dipole interactions between the atoms in the system, the strength of which is determined by $V_{ij}$.
$\sigma_{i}^{\dag}$ and $\sigma_{i}$ are the raising and lowering operators for the atom $i$, respectively, such that $\sigma_{i}^{\dag}\ket{g_i}=\ket{e_i}$ and $\sigma_{i}\ket{e_i}=\ket{g_i}$. The second term in (\ref{Master eq}) represents the dissipation via the emission of photons into the environment. The dissipation in this system has in general a collective character, determined by the form and size (relative to the single atom decay rate) of the non-diagonal elements of the matrix $\Gamma_{ij}$ \cite{Lehmberg1970}. We will elaborate further on this collective character, which leads to super- and subradiant emission, in the last section of the paper.

It is convenient to use cylindrical coordinates to represent the position $(r_i,\varphi_i,z_i)$ of the $i$-th atom, assuming that the waveguide is concentric with the $z$ axis [see Fig. \ref{fig:Setup}(b)]. Both $V_{ij}$ and $\Gamma_{ij}$ can be split into the corresponding contributions from the guided modes and unguided (or radiation) modes as $V_{ij}=V^g_{ij}+V^r_{ij}$ and $\Gamma_{ij}=\Gamma^g_{ij}+\Gamma^r_{ij}$. The contribution to the interactions due to the fundamental guided modes of the waveguide are given by
\begin{eqnarray*}
  V^g_{ij}&=& -\mathrm{i}\pi \sum_{fl}\mathrm{sgn}(fz_{ij})G_{\omega_0 fli}G^*_{\omega_0 flj}\\
\Gamma^g_{ij} &=& 2\pi\sum_{fl}G_{\omega_0 fli}G^*_{\omega_0 flj}.
\end{eqnarray*}
Here, $z_{ij}=z_i-z_j$, $f=\pm1$ represents the propagation direction of the guided mode in the $\pm z$ direction and $l=\pm1$ its polarisation in the clockwise or anticlockwise direction, respectively. The Green's function $G_{\omega_0 fli}$ couples atom $i$ to the guided mode of frequency $\omega_0$, with $G^*_{\omega_0 fli}$ denoting its complex conjugate. In the case of a nanofiber, this function reads
\begin{equation}\label{Gs}
    G_{\omega_0 fli}\!=\!\sqrt{\frac{\omega_0 \beta'}{4\pi \hbar \epsilon_0}}\left[\boldsymbol{d}_i\cdot \boldsymbol{e}^{(\omega_0 fl)}(r)\right]e^{\mathrm{i}(f\beta z_i + l\varphi_i)}
\end{equation}
where $\epsilon_0$ is the vacuum permittivity, $\boldsymbol{d}_i=\left(d_r^{\,i}, d_{\varphi}^{\,i}, d_z^{\,i} \right)^T$ is the transition dipole moment of atom $i$, $\boldsymbol{e}^{(\omega fl)}(r)=\left(e_r, -le_{\varphi}\,, fe_z \right)^T$ is the electric field profile function of the guided mode evaluated at the distance $r$ from the atom chain to the $z$ axis (the center of the nanofiber), and $\beta$ and $\beta'$ are the longitudinal propagation constant within the waveguide and its derivative, respectively.

The specific choice of orientation and position of the atomic dipole moments with respect to the nanofiber and with respect to each other gives rise to different coherent and incoherent interactions. Moreover, the atomic dipole moment orientation determines whether the emission into the nanofiber is directional or not, i.e., whether the emission of photons from a single atom occurs in a preferential direction (i.e., to the right- or the left-propagating modes of the nanofiber). For simplicity, we will consider here the situation where all dipole moments are perpendicular to the nanofiber, i.e. $d_r^i=d\equiv\left|\mathbf{d}_i\right|$, which leads to a symmetric emission to the right and to the left. Moreover, all atoms are placed on a line, such that $\varphi_{ij}=\varphi_i-\varphi_j=0$ for all $i,j=1,\dots M$. The coefficients representing the interactions read here
\begin{eqnarray}
\label{eq:Vij}
        V_{ij}^g&=&\frac{\gamma}{2}\mathrm{sgn}(z_{ij})\sin(\beta z_{ij}),\\
        \label{eq:Gammaij}
        \Gamma_{ij}^g&=&\gamma\cos(\beta z_{ij}),
\end{eqnarray}
where
\begin{equation}
\gamma=\frac{2d^2\omega \beta'}{ \hbar \epsilon_0}\abs{e_r}^2    
\end{equation}
is the single atom decay rate into the waveguide. Note, that the interactions induced by these guided modes do not decay with the distance. In this paper, we will be only considering the interactions mediated via these guided modes. This is a particularly good approximation when the distances $a$ and $b$ are comparable or larger than the wavelength $\lambda_0=2\pi c/\omega_0$ of the $\ket{g}\to\ket{e}$ transition as, contrary to the guided case,  both $V^r_{ij}$ and $\Gamma^r_{ij}$ decay as $\left(\lambda_0/z\right)^3$, with $z$ being the distance between two atoms \cite{Asenjo-Garcia2017,Zhang2020}.

We start by considering the Hamiltonian (\ref{eq:Hprime}) and consider the effects of dissipation in a later section. Hamiltonian (\ref{eq:Hprime}) conserves the number of excitations in the system. Constraining our analysis to the single-excitation sector, it can be mapped via a Jordan-Wigner transformation into a quadratic Hamiltonian for spinless fermions
\begin{equation}\label{eq:extendedSSH}
    H=\hbar\sum_{i<j}J_{|i-j|} c_i^\dag c_j+\text{h.c.}
\end{equation}
where $J_{|i-j|}=V_{ij}^g$ and $c_j$ and $c_j^\dag$ are the annihilation and creation operators of a spinless fermion on site $j$, respectively. Due to the bipartite form of the lattice, we can describe this system as a one-dimensional array formed by $M$ sites distributed over $M/2$ unit cells, each hosting two sites belonging to two distinct sublattices, A and B [see Fig. \ref{fig:Setup}(a)]. Hopping rates differ when the two sites involved belong to the same or different sublattices (i.e. $\abs{i-j}$ even and odd, respectively). Moreover, in the latter case we distinguish the case when $i$ is odd or even, i.e., when the hopping occurs between sites separated by $p-1$ or $p$ unit cells, respectively. In summary, we denote the hopping rates [see Fig. \ref{fig:Setup}(a)] as
\begin{equation}\label{coeff notation}
    J_{|i-j|}=\left\{\begin{array}{cc}
        J_{2p} & \text{$\abs{i-j}$ even}\\
        J_{2p-1} & \text{$\abs{i-j}$ odd; $i$ even}\\
        J'_{2p-1} & \text{$\abs{i-j}$ odd; $i$ odd}
    \end{array}\right.
\end{equation}
with $p=1,\dots M/2$ \cite{perez-gonzalez2019}. Hence, the Hamiltonian can be rewritten as
\begin{eqnarray*}
    &&H=\hbar\left\{\sum_{q=1}^{M/2}\sum_{p=1}^{M/2-q}\!\left[J_{2p-1} b_{q}^\dag a_{q+p}+J'_{2p-1} a_{q}^\dag b_{q+p-1}\right. \right.\\
    &&\left.\left.+J_{2p}\left(a_q^\dag a_{p+q}+b_q^\dag b_{p+q}\right)\right]\!+J'_{M-1} a^\dagger_1 b_{M/2}\right\}+\text{h.c.},
\end{eqnarray*}
where we have introduced $a_q=c_{2i-1}$ and $b_q=c_{2i}$ as the annihilation operators of a fermion in the $q$-th unit cell in sublattices A and B, respectively. Note, that when all hopping rates are set to zero except $J_1$ and $J'_1$, this Hamiltonian describes the well-known SSH model \cite{su1979}. In our case, considering the form of (\ref{eq:Vij}), we obtain
\begin{align}
    \label{hop even}
    &J_{2p}=\frac{\gamma}{2}\sin{(\beta ap)}, &&\\ 
    \label{hop odd1}
    &J'_{2p-1}=\frac{\gamma}{2}\sin{\left[\beta((p-1)a+b)\right]},    &&\\ 
    \label{hop odd2}
    &J_{2p-1}=\frac{\gamma}{2}\sin{\left[\beta(ap-b)\right]}.  &&
\end{align}
Hence, in this system the values of all the hopping rates are real and restricted to the interval $[-\gamma/2,\gamma/2]$.

\section{Bulk properties}

We start by analyzing the bulk properties of Hamiltonian (\ref{eq:extendedSSH}) assuming periodic boundary conditions. Here, the bulk Hamiltonian becomes a real symmetric block circulant matrix ($2\times 2$ blocks), allowing us to diagonalize the bulk Hamiltonian in the basis of Bloch states for the external degrees of freedom, i.e. cell index, as
\begin{equation}
    H_\mathrm{bulk}=\hbar\sum_{k}\Psi_k^\dag h(k) \Psi_k,
\end{equation}
with $k\in\left[0,\frac{2\pi}{a}\right]$. Here, we have introduced the eigenvectors 
\begin{equation}
    \Psi_k=\begin{pmatrix}
    \tilde{a}_k\\
    \tilde{b}_k
    \end{pmatrix}=\frac{1}{\sqrt{\floor{M/4}}}\sum_{q=1}^{\floor{M/4}} \mathrm{e}^{\mathrm{i}kqa}\begin{pmatrix}
    a_q\\
    b_q
    \end{pmatrix}.
\end{equation}
It is convenient to write the bulk Hamiltonian $h(k)$ as
\begin{equation}
    h(k)= h_{e}(k)+h_{o}(k)
\end{equation}
where we have separated the parts containing the intra-sublattice (even, diagonal) hopping coefficients
\begin{equation}
 h_{e}(k)= \begin{pmatrix}
        n_0(k) & 0 \\ 
        0 & n_0(k)
    \end{pmatrix}  
\end{equation}
with
\begin{equation}
    n_0(k)=2\sum_{p=1}^{\floor{M/4}}
        J_{2p}\cos{kpa},
\end{equation}
and the inter-sublattice (odd, off-diagonal) hoppings,
\begin{equation}
 h_{o}(k)= \begin{pmatrix}
        0 & n(k) \\ 
        n^*(k) & 0
    \end{pmatrix}
\end{equation}
with
\begin{equation}
    n(k)\!=\!\!\sum_{p=1}^{\floor{M/4}}\!
        \left[J_{2p-1}\mathrm{e}^{\mathrm{i}kpa}+J'_{2p-1}\mathrm{e}^{-\mathrm{i}k(p-1)a}\right].
\end{equation}
The two energy bands are then determined by $E^{\pm}(k) = n_{0}(k) \pm \,\abs{n(k)}$.

The relevant symmetries to determine the topology of this system according to the tenfold way classification \cite{ryu2010} are time-reversal, particle-hole, and chiral symmetry. These are determined in our case by the operators $T={\cal K}$, $C=\sigma_z {\cal K}$ and $S=T\cdot C=\sigma_z$, respectively, where ${\cal K}$ is the conjugation operator and $\sigma_z$ is the usual spin-$1/2$ Pauli matrix. In general, for real hopping coefficients, $h_{o}(k)$ exhibits time-reversal symmetry $Th_{o}(k)T^{-1}=h_{o}(-k)$, particle-hole symmetry $Ch_{o}(k)C^{-1}=-h_{o}(-k)$ and chiral symmetry $Sh_{o}(k)S^{-1}=-h_{o}(k)$. However, $h_{e}(k)$ only possesses time-reversal symmetry and, most importantly, its presence makes the full Hamiltonian break chiral symmetry. This system is therefore to be attributed to the AI symmetry class, for which there is no topological invariant in one dimension. On the other hand, when $n_{0}(k)=0$ (e.g., when the all even hoppings $J_{2p}=0\,\,\, \forall p$) the system is attributed to the BDI symmetry class, which has a $\mathbb{Z}$-type topological invariant in one dimension. This is, for example, the case for the usual SSH model in one dimension. The chiral symmetry is also reflected on the eigenenergies of the system, which lead to a spectrum that is symmetric about $E=0$ when $n_0(k)=0$. In real space, the consequence of chiral symmetry is that if a pair of degenerate eigenstates have zero energy, each eigenstate can be rewritten such that they have support only on sublattice A and B, respectively. Conversely, if an eigenstate has support only on one sublattice, its energy must be zero \cite{Chen2020}. In the following, unless stated otherwise, we will work on the assumption that chiral symmetry is indeed preserved.

The intrinsic geometric phase of the eigenstates of $h(k)$ is given by the variation of the argument of $n(k)=n_{x}(k) +in_{y}(k)$ through the First Brillouin Zone (FBZ). The winding number of $n(k)$ can be directly mapped to the Berry phase and thus be used to classify the intrinsic geometry, and hence topology, of the system. The winding number of the system can be found as
\begin{equation}\label{Winding Number}
    \nu=\frac{1}{2\pi}\Delta \phi(k),
\end{equation}
where $\Delta \phi(k)$ is the variation of the argument of $n(k)$ through the FBZ. 
\begin{figure}
    \includegraphics[width =\columnwidth]{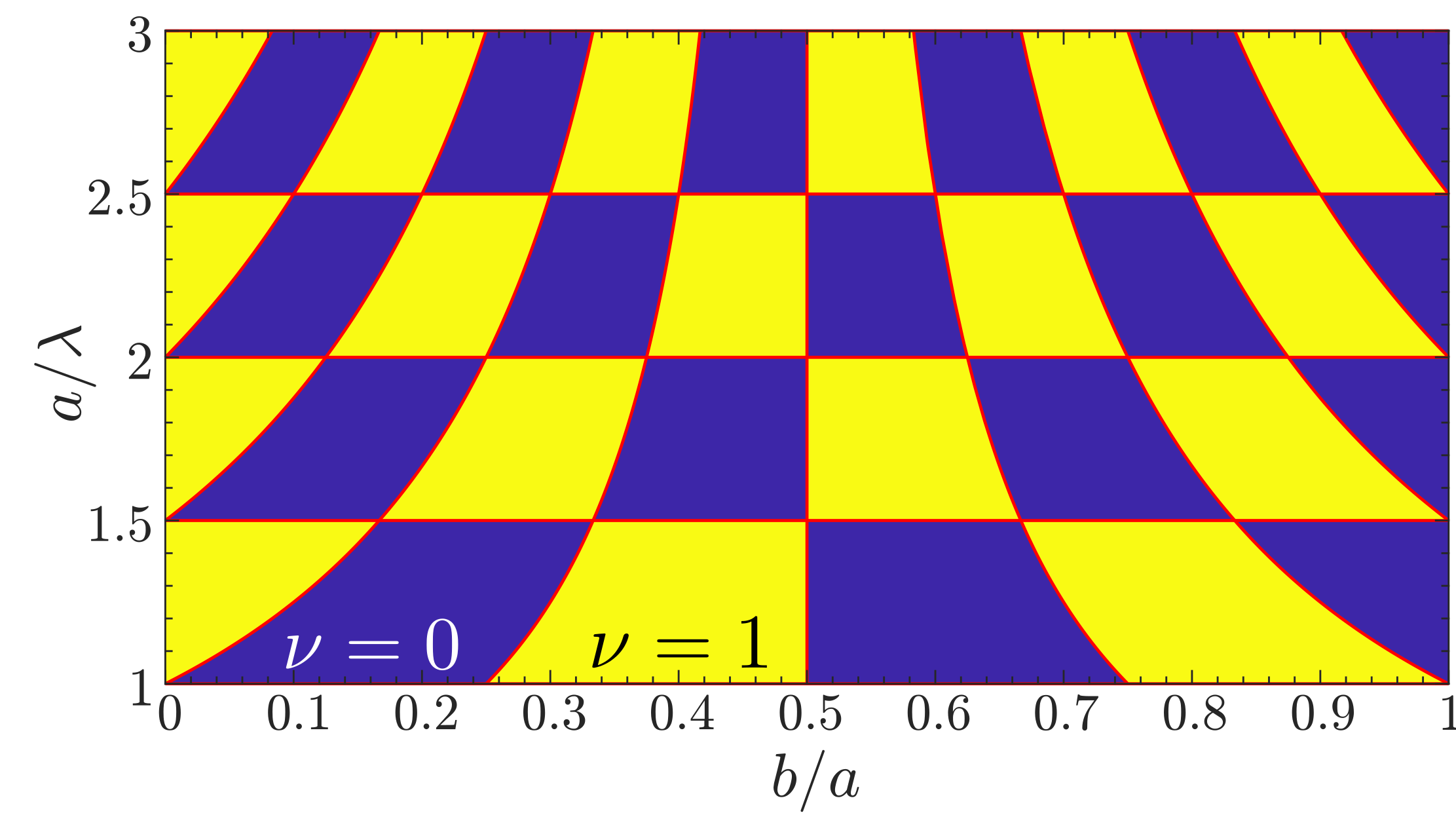}
    \caption{\textbf{Winding number.} The winding number $\nu$, depicted here as a function of $a/\lambda$ and $b/a$, takes only the values 0 and 1. The red solid lines represent the analytical predictions (\ref{analytic phases b}) and (\ref{analytic phases a}) for the phase boundaries. } 
    \label{fig:Winding}
\end{figure}

Due to the presence of all-to-all hopping rates, the function $n(k)$ depends on the number of atoms $M$ even when considering periodic boundary conditions. However, we find when evaluating the winding number numerically that it only takes the values $\nu=0$ and $1$ (see Fig. \ref{fig:Winding}) and that it does not vary with the system size. The values of $b$ and $a$ for a fixed $\beta$ at which the system changes from a topologically trivial to non-trivial phase and vice versa can be, moreover, obtained analytically by analyzing the symmetries of $n(k)$: as a consequence of time-reversal symmetry, $n_x(k)$ and $n_y(k)$ are symmetric and antisymmetric, respectively, with respect to $k=\pi/a$. Hence, the closed path of $n(k)$ in complex space within the FBZ is symmetric about the real axis. In order to determine the winding number, we need to verify whether the closed path of $n(k)$ encloses the origin ($\nu=1$) or not ($\nu=0$). Since at the time reversal invariant momenta $k=0$ and $\pi/a$, $n(k)$ crosses the real axis, i.e., $n_y(k)=0$, it is enough to determine whether there is a change of sign in $n_x$ at the two momenta. In particular, if $\text{sgn}(\frac{n_{x}(\pi/a)}{n_{x}(0)})=1$, then $\nu=0$, and if $\text{sgn}(\frac{n_{x}(\pi/a)}{n_{x}(0)})=-1$, then $\nu=1$, such that
\begin{equation}\label{analytic nu}
    \nu=\frac{1}{2}\left[1 -\text{sgn}\left(\frac{n_{x}(\pi/a)}{n_{x}(0)}\right)\right].
\end{equation}
We can determine the phase boundaries analytically using that
\begin{equation}\label{sgn ratio}
    \frac{n_{x}(\pi/a)}{n_{x}(0)}\propto\tan\left[\beta\left(b-\frac{a}{2}\right)\right]\tan\left[\frac{\beta a}{2}\right].
\end{equation}
Hence, one easily extracts that the winding number changes at the values
\begin{equation}\label{analytic phases a}
    \frac{a}{\lambda}=\frac{m}{2}\quad \quad \text{where } m\in\mathbb{Z},
\end{equation}
where $\lambda=\frac{2\pi}{\beta}$ is the wavelength of the light propagating in the waveguide. Note, that due to the presence of all-to-all interactions, the gap between the two energy bands also closes whenever $a/\lambda=n/M$ for any $n\in\mathbb{Z}$. However, only when condition (\ref{analytic phases a}) is satisfied is this gap closing accompanied by a change in the topological invariant of the system. Finally, fixing $a$, we also find a change from $\nu=0$ to $\nu=1$ and viceversa when
\begin{equation}\label{analytic phases b}
    \frac{b}{\lambda} = \frac{1}{2}\frac{a}{\lambda}+\frac{n}{4}, \quad \quad \text{with } n\in\mathbb{Z}.
\end{equation}

Note, that when only $J_1$ and $J'_1$ are different from zero, the Hamiltonian (\ref{eq:extendedSSH}) is precisely the one describing the SSH model, for which $\nu=1$ when $\abs{J_{1}}>\abs{J'_{1}}$ while $\nu=0$ when $\abs{J_{1}}<\abs{J'_{1}}$ \cite{Batra2020}. Interestingly, we find that the winding number for our fully connected model (Fig. \ref{fig:Winding}) coincides with the one of the SSH model, i.e., the topological invariant of this system does not change as the connectivity of the problem is increased from nearest neighbors to all-to-all interactions.

\section{Weakened bulk-boundary correspondence}

In this section, we study not only the bulk but also the boundaries of the system, i.e., we assume open boundary conditions. For a Hamiltonian with short-ranged hopping, the bulk-boundary correspondence establishes that, in a parameter regime where winding number is $\nu=1$, the system will host one pair of energy eigenstates inside the insulating gap between conduction and valence bands. These states are expected to be exponentially localised on the edges of the chain \cite{Rhim2018}. For a semi-infinite chain, these eigenstates have exactly zero energy, hence the name `zero-energy edge states'. The bulk-boundary correspondence gets weakened in the presence of long-ranged hopping due to the correlations that are built between the bulk and the edges of the system, giving rise to massive edge states, i.e., edge states with energy $\varepsilon$ different from zero (so-called mass gap), which may decay from the boundary into the bulk algebraically instead of exponentially \cite{lepori2017}.

We illustrate here the weakening of the bulk-boundary correspondence for our system with all-to-all interactions, when chiral symmetry is conserved (all even couplings $J_{2p}=0$). The Hamiltonian is given by 
\begin{eqnarray*}
    H=\hbar&&\left\{\sum_{q=1}^{M/2}\sum_{p=1}^{M/2-q}\left[J_{2p-1} b_{q}^\dag a_{q+p}+J'_{2p-1} a_{q}^\dag b_{q+p-1}\right] \right.\\
    &&\left.+J'_{M-1} a^\dagger_1 b_{M/2}\right\}+\text{h.c.}.
\end{eqnarray*}

\begin{figure*}
\centering
    \includegraphics[width =0.9\textwidth]{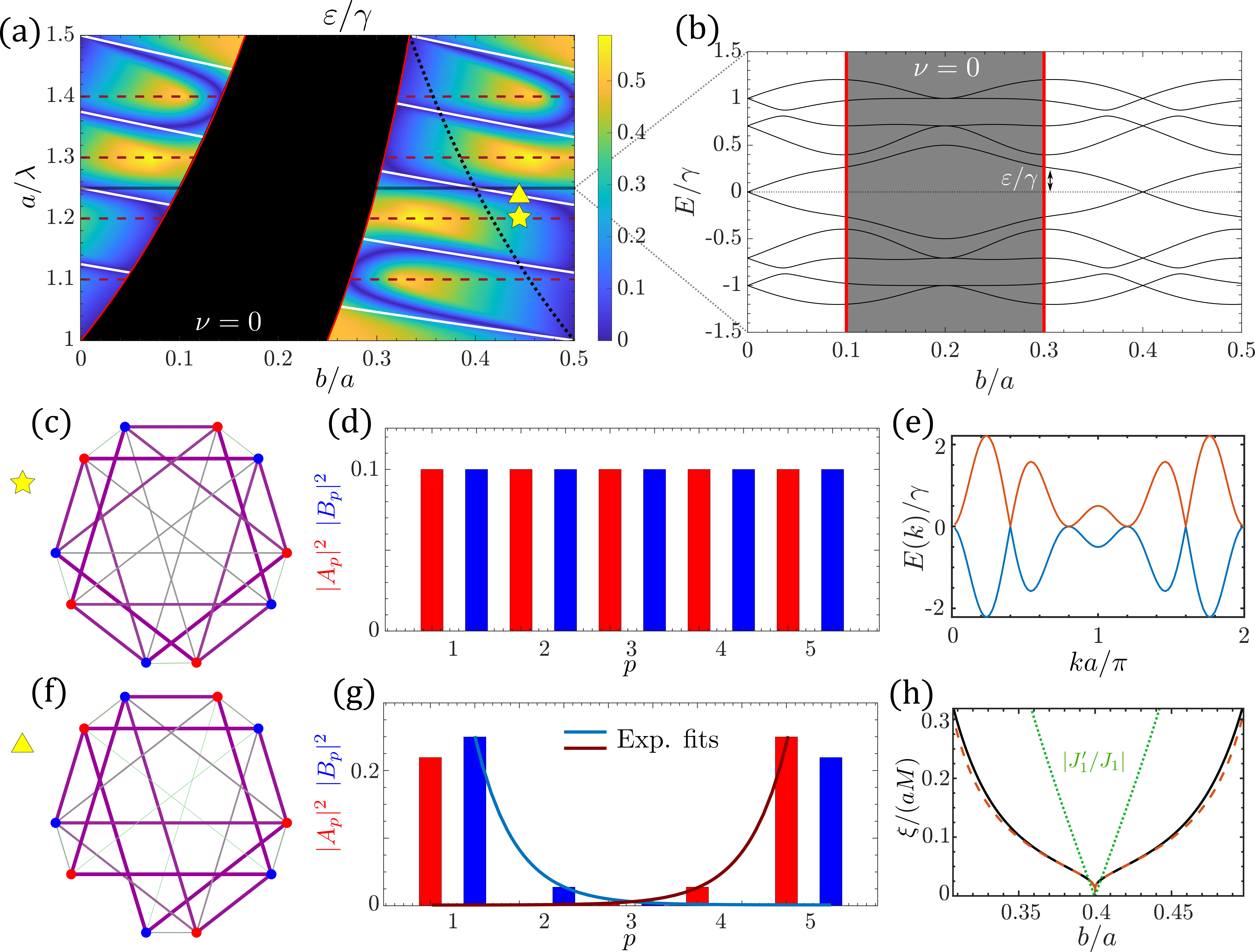}
    \caption{\textbf{Weakened Bulk-Boundary correspondence.} \textbf{(a)}: Mass gap $\varepsilon$ for a system of $M=10$ sites as a function of $a/\lambda$ and $b/a$. The black area represents the region where $\nu=0$. The horizontal dashed, solid white, and black dotted lines indicate where $|J'_{M-1}|=|J_1|$, $J'_{M-1}=0$, and $J'_{1}=0$, respectively. \textbf{(b):} Spectrum at $a/\lambda=1.25$ varying $b/a$ (black lines are the eigenenergies $\varepsilon_n$ with $n=1,\dots,10$). The mass gap is exactly zero at $b/a=0.4$. \textbf{(c):} Sketch of the (chiral) system's couplings at $a/\lambda=1.2$ and $b/a=0.4444$ ($|J'_{M-1}|=|J_1|$). \textbf{(d):} The system is translationally symmetric, as reflected in the population of one of the eigenstates, and \textbf{(e):} the gap in $k$-space is closed. \textbf{(f):} Sketch of the (chiral) system's couplings at $a/\lambda=1.2375$ and $b/a=0.4444$ ($J'_{M-1}=0$). \textbf{(g):} Near zero-energy states are exponentially localized to the edges of the system. \textbf{(h):} Localization length along one of the $J'_{M-1}=0$ lines. The black solid line represents the length obtained from a fitting while the red dashed line is the analytical approximation given by (\ref{eq:loc_approx}). The green dotted line is the ratio $|J'_1/J_1|$.}
    \label{fig:BB}
\end{figure*}

In Figure \ref{fig:BB}(a) we show the mass gap $\varepsilon$ for a system of $M=10$ sites as a function of $a/\lambda$ and $b/a$. In order to calculate the mass gap, we obtain the eigenenergies $\varepsilon_n$ with $n=1,\dots, M$ of the Hamiltonian $H$ for each value of $a/\lambda$ and $b/\lambda$. Due to the imposed chiral symmetry, the resulting spectrum is symmetric around $E=0$ [see, e.g., Fig. \ref{fig:BB}(b)]. The mass gap is thus calculated as
\makeatletter\onecolumngrid@push\makeatother
\clearpage
\twocolumngrid\makeatother
$\varepsilon=\left|\varepsilon_{M/2+1}-\varepsilon_{M/2}\right|/2$. Note, that the mass gap becomes simply half of the gap between the two bulk bands in the topologically trivial area, i.e. where $\nu=0$ (c.f. Fig. \ref{fig:Winding}). 

Since in this system the interactions are all-to-all, the specific values of the couplings (determined in turn by $a/\lambda$ and $b/\lambda$) give rise to a kaleidoscope of regimes with different effective geometries and edge state properties. In the following, we will analyze in detail some of these parameter regimes.

\subsection{Flat states ($|J'_{M-1}|=|J_1|$)}

In this case, the first particle can hop to its nearest neighbor and to the last site of the chain with the same hopping rate. Moreover, due to the specific form of the rates (\ref{hop odd1}-\ref{hop odd2}), imposing that $|J'_{M-1}|=|J_1|$ automatically implies that $|J'_{M-2p+1}|=|J_{2p-1}|$ for all $p$ [see Fig. \ref{fig:BB}(c)]. Here, not only the bulk but also the finite open system are translationally symmetric. Since all cells are equivalent, the concept of boundary becomes ill-defined. As a consequence, as one can observe in Fig. \ref{fig:BB}(d), all eigenstates of the Hamiltonian are such that all sites are uniformly occupied, reflecting the translational symmetry. Note, that the condition $|J'_{M-1}|=|J_1|$ is satisfied when $a/\lambda=m/M$ with $m\in\mathbb{N}$. As we pointed out in the previous section, here the bulk gap in momentum space closes [see Fig. \ref{fig:BB}(e)]. Hence, even though this gap closing does not come accompanied by a change in the topological invariant, specifically in this parameter regime [horizontal dashed lines in Fig. \ref{fig:BB}(a)] the winding number is not well defined and hence the flat states are not topological edge states.
    
\subsection{Localized states ($J'_{M-1}=0$)}

The translational symmetry is broken whenever $|J'_{M-1}|\neq|J_1|$, as one can again distinguish the boundaries from the bulk. The extreme case here occurs when $J'_{M-1}=0$, i.e., when the first and the last atom in the chain are not coupled to each other [see Fig. \ref{fig:BB}(f)]. This occurs whenever the parameters satisfy
\begin{equation}
    \frac{b}{\lambda}=\frac{M-2}{2}\frac{a}{\lambda}+\frac{m}{2}\quad m\in\mathbb{N}.
\end{equation}
In the topologically non-trivial region and along this line in parameter space [solid white lines in Fig. \ref{fig:BB}(a)], one finds two approximately zero energy [mass gap $\varepsilon\approx 0$] eigenstates. 

While usual topologically protected edge states are exponentially localized on one of the two boundaries, we find here that the population is concentrated on both boundaries of the chain [cf. Fig. \ref{fig:BB}(g)], such that the occupation probability of the sites in the $p$-th cell in sublattice A and B, respectively, can be written as
\begin{eqnarray*}
    |A_{M/2-p+1}|^2&=&|A_{M/2}|^2 e^{-\frac{pa}{\xi}},    \\
    |B_p|^2&=&|B_1|^2 e^{-\frac{pa}{\xi}},
\end{eqnarray*}
for $p=2,\dots,M/2-1$, and $|A_1|^2=|B_{M/2}|^2$. Here, we have introduced $\xi$ as the localization length of the edge state for each sublattice. As we can observe in Fig. \ref{fig:BB}(h), the localization length is smaller (i.e. edge states more localized in the edges of the lattice) the smaller the ratio $J'_1/J_1$. One can obtain that the localization length is approximately given by
\begin{equation}\label{eq:loc_approx}
    \xi\approx -\frac{a}{\log{\left|\frac{J'_1}{J'_3}\right|}}.
\end{equation}
As one can see in Fig. \ref{fig:BB}(h), this simple expression yields indeed an excellent approximation to the localization length, particularly for small values of the ratio $|J'_1/J_1|$. Note also that the edge states have support of both sublattices A and B, instead of only on one as it is the case in the SSH model. One recovers this feature only when the energy of the localized state is exactly zero.

The edge states become completely localized on the edges when $J'_1=0$. It can be easily shown that this occurs when 
\begin{equation}\label{eq:ZE_cond}
    \frac{a}{\lambda}=\frac{m}{M-2},\quad \frac{b}{\lambda}=\frac{m'}{2},
\end{equation}
with $m$ and $m'$ being natural numbers. Here, the edge states can be analytically obtained, and they read
\begin{equation}\label{eq:fully_loc_edge}
\!\!\!\!\left|\Psi_\pm\right>\!=\!\frac{1}{2}\!\left[a_1^\dag\!\pm\! b_{1}^\dag+(-1)^n \!\left(a_{M/2}^\dag\!\pm\! b_{M/2}^\dag\right)\right]\!\left|0\right>\!,
\end{equation}
with $n=\lfloor M/4\rfloor$, and $\left|0\right>$ being the state with no fermions. Note, moreover, that at this point these edge states have exact zero energy, and that the spectrum becomes doubly degenerate [see Fig. \ref{fig:BB}(b)]. This structure is reminiscent of what occurs in the presence of a so-called strong zero mode \cite{Fendley2016,Else2017,Kemp2017,Vasiloiu2018,Vasiloiu2019,Vasiloiu2022,Klobas2022}. Indeed, here there exists an operator $\Psi=\sum_p \mathrm{i}(-1)^p\left(a^\dag_p b_p-b^\dag_p a_p\right)$ that anticommutes with a symmetry of the Hamiltonian ${\cal D}=\sum_p\left(a^\dag_p b_p+b^\dag_p a_p\right)$. Due to the long-range character of the hopping rates, though, in our case the strong zero mode $\Psi$ is not concentrated on the edges of the lattice, as it is usually the case considering systems with short range interactions. Moreover, the Hamiltonian exactly commutes with $\Psi$ for all system sizes (cf. \cite{Klobas2022}), instead of doing so only in the thermodynamic limit.

\subsection{Robustness against disorder}

\begin{figure}
\centering
    \includegraphics[width =\columnwidth]{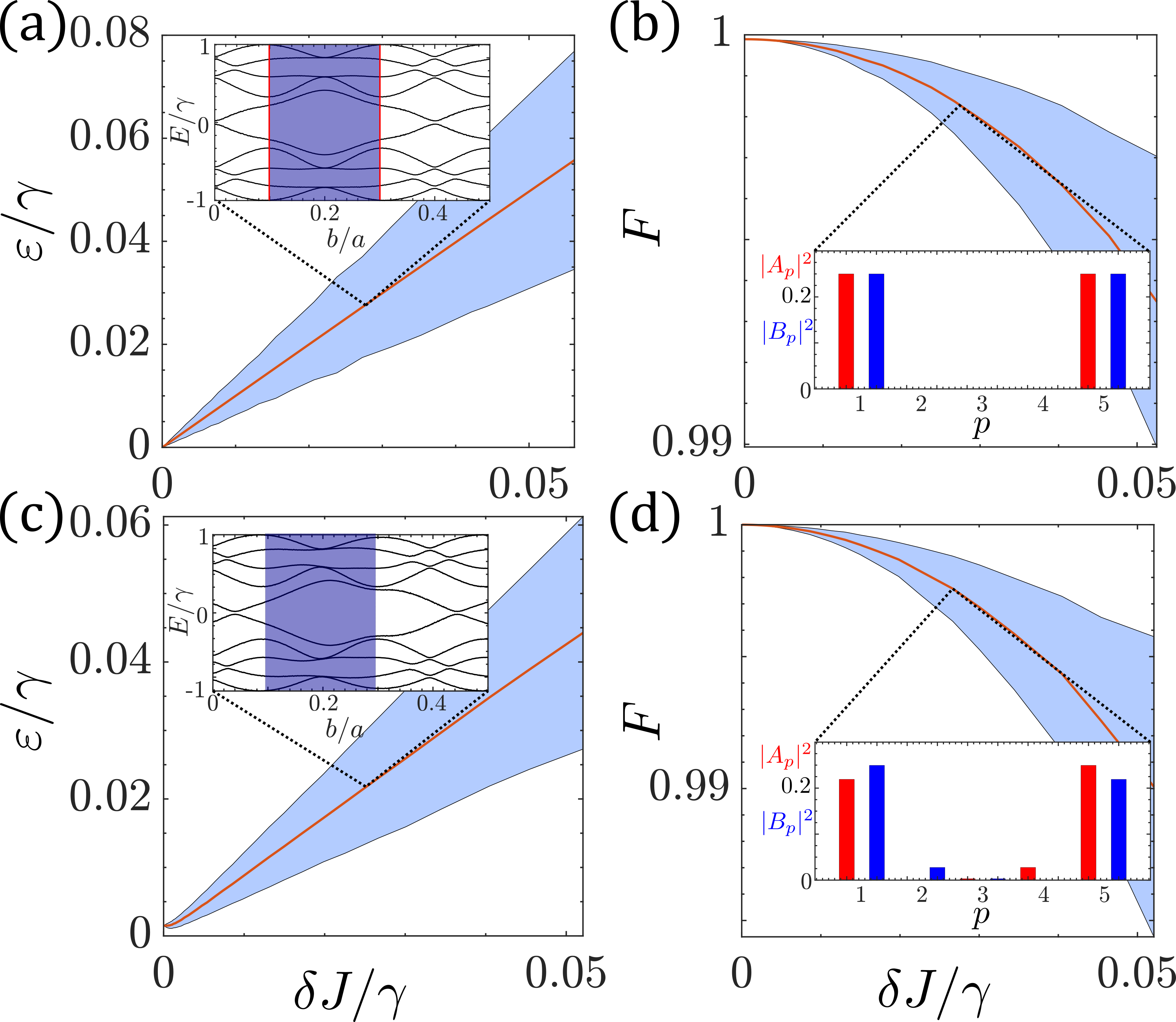}
    \caption{\textbf{Robustness against local disorder.} \textbf{(a)} and \textbf{(c):} Mass gap $\varepsilon$ for a system of $M=10$ sites as a function of the average fluctuations of the coupling constants $\delta J$.  \textbf{(b)} and \textbf{(d):} Fidelity $F$ of the edge state in the presence of disorder with respect to the corresponding state in the absence of disorder. In all figures, the red line represents the average value and the blue shaded area the standard deviation after 5000 realizations of disorder. The parameters are $a/\lambda=1.25$ and $b/a=0.4$ for \textbf{(a)} and \textbf{(b)} and $a/\lambda=1.2375$ and $b/a=0.4444$ for \textbf{(c)} and \textbf{(d)}.}
    \label{fig:Disorder}
\end{figure}

Topologically protected edge states are robust against local disorder. In order to test whether the edge states described above are indeed robust, we introduce disorder as an uncertainty in the position of the atoms in the chain. We do this by letting the positions of the atoms $z_i$ in the chain have a Gaussian distribution around their non-disordered positions $z^{(0)}_{i}$, i.e.
\begin{equation}\label{Disorder}
    z_{i} \rightarrow z^{(0)}_{i} +\sigma R,
\end{equation}
where $R$ is a number randomly generated from a normal distribution centred around zero and $\sigma$ is the standard deviation of the normal distribution that is used to control the level of disorder applied. For each realization of disorder, we calculate the mass gap $\varepsilon_\sigma$ of the edge state $\left|\Psi_{\sigma}\right>$, and the fidelity
\begin{equation}
    F_\sigma=\left|\left<\Psi_{\sigma}|\Psi_{+}\right>\right|^2,
\end{equation}
between the measured edge state and the state with zero disorder. After many realizations, we extract the corresponding average fluctuations of the hoppings, $J\to J+\delta J$, and the average mass gap $\varepsilon$ and fidelity $F$, shown in Fig. \ref{fig:Disorder} (a) and (b), respectively, for the fully localized case, where both $J'_1$ and $J'_{M-1}$ are equal to zero [where in the absence of disorder $\varepsilon=0$ and the edge states are exactly given by (\ref{eq:fully_loc_edge})]. As one can observe, as we increase the average fluctuations up to more than a $10\%$ (note that the maximum value $\delta J$ could take is $\gamma/2$) the mass gap increases linearly with the disorder, remaining much smaller than the gap between the bulk bands in all cases. Moreover, the fidelity of the edge state with respect to the no-disorder case remains above $0.99$. In Fig. \ref{fig:Disorder} (c) and (d) we show the same results for a parameter regime where $J'_{M-1}=0$ but $J'_1\neq 0$ [cf. Fig. \ref{fig:BB}(f-g)]. One can easily observe that the deviation of the mass gap and fidelity from their no-disorder values remain small. Hence, we have established that the edge states along the full line $J'_{M-1}=0$ are indeed extremely robust against external disorder. Finally, we have verified numerically that both the overall structure of the edge states and their robustness remain unchanged for system sizes ranging from $M=6$ to $M=500$ atoms. 

\section{Edge states with $J_{2p}\neq 0$}

Up until now we have studied a system where the chiral symmetry is not broken, i.e., where the intra-sublattice hoppings $J_{2p}$ have been put to zero. In a real atom-waveguide system, these cannot be generally put to zero unless $a/\lambda=m/2$ [see Eq. (\ref{hop even})]. Unfortunately, this regime is also precisely where the bulk gap closes (horizontal lines in Fig. \ref{fig:Winding}), and hence no topological invariant can be defined. In this section, we will explore a situation where, even though the chiral symmetry is not strictly conserved, states with energy close to zero appear in the gap between the bulk bands which, moreover, are localized to the edges of the chain and are robust to local disorder.

\begin{figure}
\centering
    \includegraphics[width =\columnwidth]{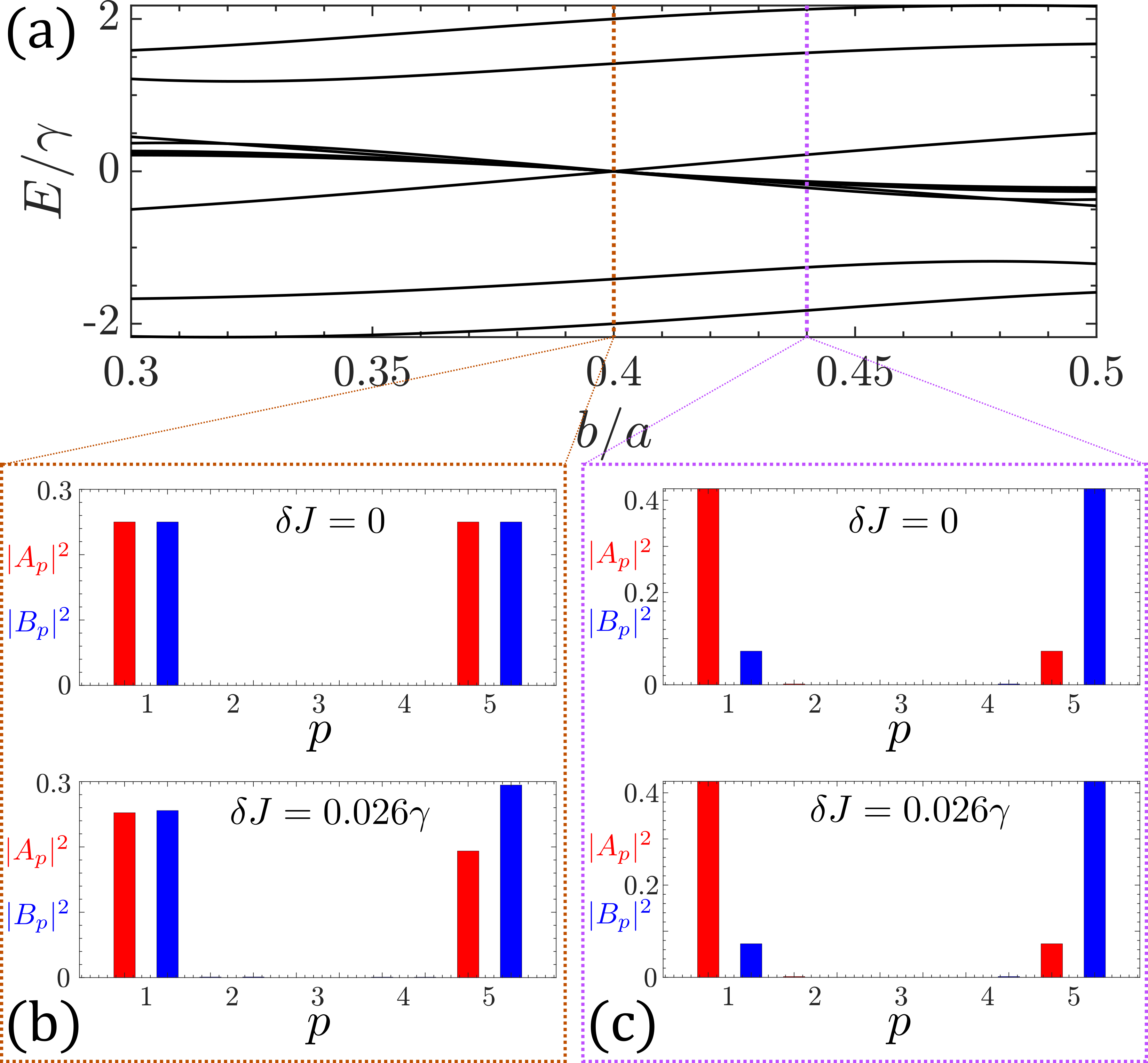}
    \caption{\textbf{Non-chiral edge states.} \textbf{(a):} Spectrum for a system of $M=10$ atoms with $a/\lambda=1.25$. \textbf{(b):} At $b/a=0.4$, the spectrum is symmetric around $E=0$ and some of the zero-energy eigenstates are strongly localized to the edges. \textbf{(c):} At $b/a=0.44$ the degeneracy is lifted, but the edge states are still present. The lower panels in \textbf{(b)} and \textbf{(c)} represent the edge state in the presence of disorder, averaged over 5000 realizations.}
    \label{fig:WithJ2}
\end{figure}

To find these states, we consider the number of atoms to be even but not a multiple of four, i.e. $M=4n+2$ for $n\in \mathbb{N}$ such that, at the point determined by condition (\ref{eq:ZE_cond}), one can make exactly half of the even couplings zero, i.e.
\begin{equation}
    J_{2p}=\left\{\begin{array}{cc}
    0     &  \text{if}\quad p\,\text{even}\\
    \frac{\gamma}{2}(-1)^{\frac{p-1}{2}}      & \text{if}\quad p\,\text{odd}.
    \end{array}\right.
\end{equation}
Moreover, as it can be observed in Fig. \ref{fig:WithJ2}(a), the spectrum becomes symmetric around $E=0$. Here, the eigenstates of the Hamiltonian can be obtained analytically. In particular, we find that there is a manifold of $M/2+1$ eigenstates with zero energy. Some of these eigenstates are indeed localized in the edge, such as
\begin{eqnarray}
    &&\!\!\!\!\!\left|S_\pm\right>=\!\frac{1}{2}\!\left[a_1^\dag\!+\!b_{1}^\dag\pm\left(a_{M/2}^\dag\!+\!b_{M/2}^\dag\right)\right]\!\left|0\right>\\\label{eq:A_state}
    &&\!\!\!\!\!\left|A\right>=\!\frac{1}{2}\!\left[a_1^\dag\!-\!b_{1}^\dag+(-1)^n\left(a_{M/2}^\dag\!-\!b_{M/2}^\dag\right)\right]\!\left|0\right>\!.
\end{eqnarray}
These states are, moreover, robust in the presence of local disorder, in the sense that they stay localized to the edges, not spreading into the bulk [Fig. \ref{fig:WithJ2}(b)]. However, note that the mixing with the rest of eigenstates of the quasi-degenerate manifold lead to small variations in the populations of the edge sites involved.

To avoid this mixing, one may move away from this particular symmetric point. As we can observe in Fig. \ref{fig:WithJ2}(a), both to the right and left of this point (larger and smaller $b/a$, respectively), the zero-energy degeneracy is indeed lifted. However, the eigenstates with energies now slightly different from zero are still localized on the edges and extremely robust to the addition of local disorder [see, e.g. Fig. \ref{fig:WithJ2}(c)]. Moreover, again the structure of the spectrum and the eigenstates, including the existence of robust edge states, are independent of the number of atoms in the system, $M$. Note, however, that strictly speaking these localized states are not topologically protected states, as the chiral symmetry is here broken and hence there is no topological invariant associated with the corresponding Hamiltonian.

\section{Non-equilibrium dynamics}

\begin{figure}[t!]
\centering
    \includegraphics[width =\columnwidth]{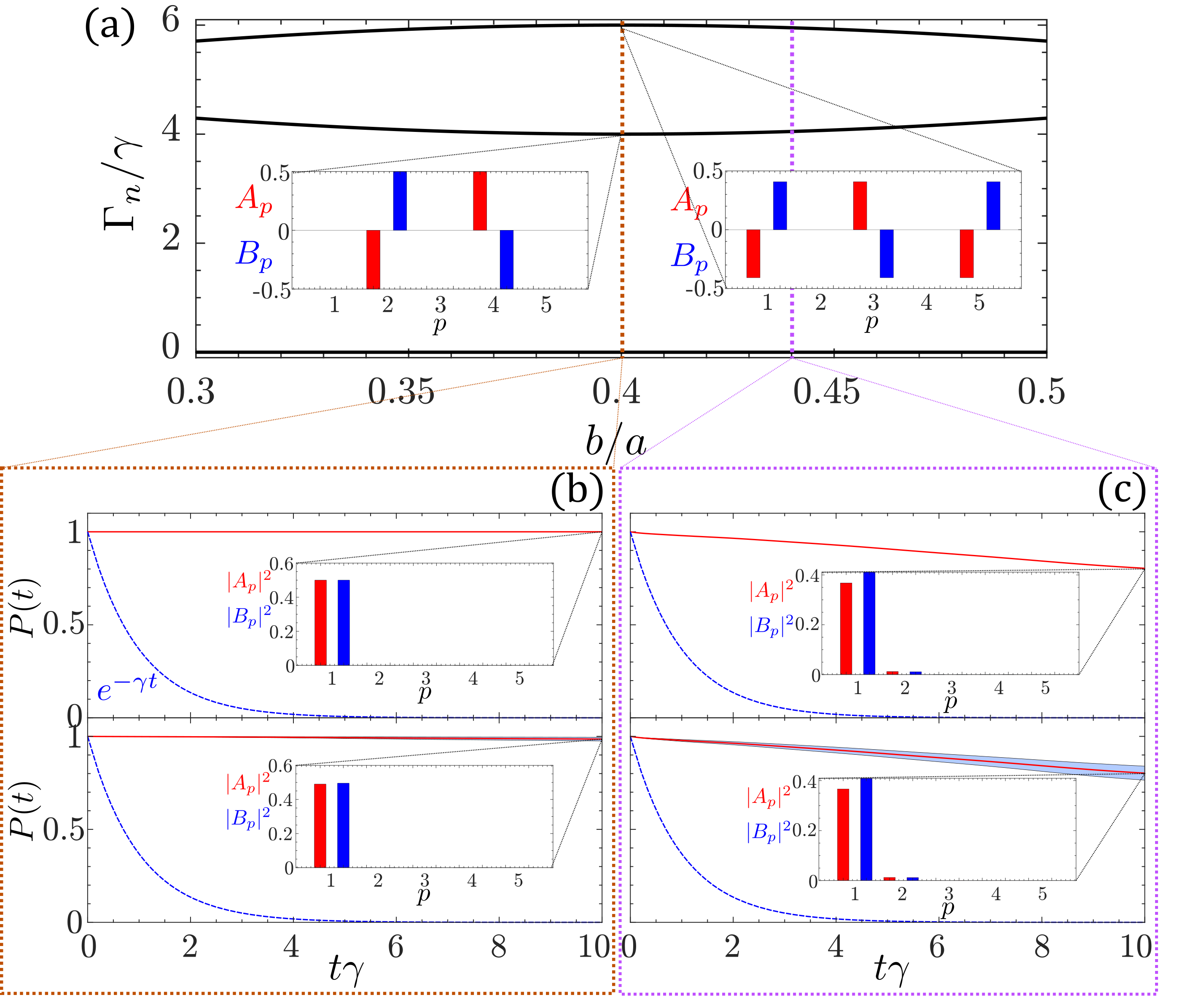}
    \caption{\textbf{Non-equilibrium dynamics.} \textbf{(a):} Collective decay rates for a system of $M=10$ atoms with $a/\lambda=1.25$. The insets show the probability amplitudes of the two superradiant states for $b/a=0.4$. \textbf{(b)} and \textbf{(c):} Survival probability $P(t)$ of a single excitation as a function of time, with the initial state being given by $\left|\Phi(0)\right>$ (see main text) for $b/a=0.4$ and $0.44$, respectively. The lower panels show the same quantities in the presence of disorder, at $\delta J=0.05\gamma$, for 5000 realizations. Each inset shows the occupation probability at the final time $t=10\gamma$. For comparison, the decay of a single atom excited next to the waveguide is depicted in all panels.}
    \label{fig:Decay}
\end{figure}

Up until now, we have studied only the static properties of the atom-waveguide system in the absence of dissipation. While this has offered us the possibility to study the topological properties of the system and demonstrate the existence of edge states, one can see from the expressions of the coherent and dissipative coupling $V^g_{ij}$ and $\Gamma^g_{ij}$, in Eqs. (\ref{eq:Vij}) and (\ref{eq:Gammaij}), respectively, that the two mechanisms are inevitably intertwined, i.e., the dissipation in general cannot be neglected. In this Section, we study the out-of-equilibrium dynamics of the system in the presence of dissipation, focusing in particular on the fate of the above found edge states.

Since the dissipation coefficients $\Gamma^g_{ij}=\gamma\cos{(\beta z_{ij})}$ do not decay with the distance between the atoms, the dissipation in this system has a collective character. We establish this by calculating the dissipation channels and corresponding decay rates $\Gamma_n$ in the single-particle sector as the eigenstates and eigenvalues of the matrix $\Gamma^g_{ij}$, respectively. Independently of the parameter regime, there are only two superradiant modes with decay rates $\Gamma_{1,2}$ much larger than $\gamma$. The rest of the states forms a subradiant manifold with zero decay rate \cite{Needham2019}, as one can observe in Fig. \ref{fig:Decay}(a) for $a/\lambda=1.25$. Along this line in parameter space, one can see that one of these two superradiant eigenstates has no support on the lattice edges, while the other one does. However, the latter has an alternating phase profile, similarly to the state $\left|A\right>$ given in Eq. (\ref{eq:A_state}). Any initial state that one chooses to excite in this system that does not have a large overlap with these two superradiant states will be, thus, subradiant.

If, moreover, we are in a parameter regime where an edge state has been predicted to exist in the previous Section, we expect an initial state with large support on the edge to remain localized in the edge during the dynamics determined by (\ref{Master eq}). This is indeed what we observe in the upper panels of Figs. \ref{fig:Decay}(b) and (c) by measuring the survival probability, $P(t)=\sum_{p=1}^{M/2}\left(|A_p(t)|^2+|B_p(t)|^2\right)$ for two values of the parameter $b/a$: the initial symmetric superposition localized state, 
\begin{equation}
\left|\Phi(t=0)\right>=\frac{1}{\sqrt{2}}\left(a_1^\dag+b_1^\dag\right)\left|0\right>
\end{equation}
(i.e., $A_1=B_1=1/\sqrt{2}$, $A_p=B_p=0$ for all $p=2,\dots M/2$) remains localized, and the excitation has an extremely long lifetime. In particular, for $b/a=0.4$ the initial state is an eigenstate of both the coherent interactions (i.e. a zero-energy edge state), and of the dissipation (belonging to the subradiant manifold). Hence, as expected, the state remains completely unchanged throughout the dynamics. In the lower panels of Figs. \ref{fig:Decay}(b) and (c) we show how this dynamics is also extremely robust against local disorder.

\section{Conclusions and Outlook}

We have studied an atom chain coupled to a waveguide and uncovered the existence of edge states, i.e. states localized in the boundary of the chain which are not only robust against the presence of local disorder, but also subradiant, i.e. they possess extremely long lifetimes.

The analysis presented here has been done in the absence of unguided or radiation modes, which has allowed us to obtain analytical results complementing the numerical ones. The contribution of the radiation modes to the dipole-dipole interactions $V_{ij}^r$ is well justified when the atoms are far from each other (compared to the wavelength $\lambda$), and are expected to represent small perturbation that does not affect the robust edge states presented here. The unguided off-diagonal components of the decay matrix, $\Gamma_{ij}^r$, can be neglected on the same grounds. However, the diagonal elements of this matrix, representing the single atom decay into the radiation continuum, can be extremely destructive for the subradiant character of the edge states. This can be remedied under the right experimental implementations such as photonic crystal waveguides \cite{Chang2018} or superconducting qubits embedded in 1D transmission lines \cite{Hoi2011}, where the decay into the guided modes can be considered predominant.

Future work will entail a closer look into the emergence of strong zero modes in our system, and whether these can also be identified in the full model including dissipation. The fate of the topological properties and edge states depicted here when the emission is not symmetric (e.g. for different atomic dipole polarizations) will also be in the focus of our future research. Finally, a natural extension of the present system is the study of the topological properties and strong zero modes in a gas of atoms coupled to a 2D waveguide structure such as a 2D photonic crystal \cite{Yu2019}.

\begin{acknowledgments}
We thank F. Carollo, J.P. Garrahan, and I. Lesanovsky for insightful comments and discussions. We acknowledge support by Open Access Publishing Fund of University of Tübingen. The authors were financially supported by the Royal Society and EPSRC [Grant Nos. DH130145 and RGF$\backslash$EA$\backslash$180021].
\end{acknowledgments}


\end{document}